\documentclass[superscriptaddress,preprintnumbers,amsmath,amssymb,aps,pre]{revtex4}
\usepackage{graphicx}
\usepackage{dcolumn}
\usepackage{bm}
\usepackage{color}
\usepackage{listings}
\usepackage[toc,page]{appendix}
\newcommand {\equalda} {\ {\raise-.5ex\hbox{$\buildrel d \over =$}}\ }
\begin{document}
\title{Seismic Electric Signals: A physical interconnection with seismicity}

\author{Panayiotis A. Varotsos}
\email{pvaro@otenet.gr} 
\affiliation{Section of Solid State Physics, Department of Physics, School of Science, National
and Kapodistrian University of Athens, Panepistimiopolis, Zografos
157 84, Athens, Greece}
\affiliation{Solid Earth Physics Institute, Department of
Physics, School of Science, National and Kapodistrian University
of Athens, Panepistimiopolis, Zografos 157 84, Athens, Greece}

\author{Nicholas V. Sarlis}
\email{nsarlis@phys.uoa.gr} 
\affiliation{Section of Solid State Physics, Department of Physics, School of Science, National
and Kapodistrian University of Athens, Panepistimiopolis, Zografos
157 84, Athens, Greece}
\affiliation{Solid Earth Physics Institute, Department of
Physics, School of Science, National and Kapodistrian University
of Athens, Panepistimiopolis, Zografos 157 84, Athens, Greece}

\author{Efthimios S. Skordas}
\email{eskordas@phys.uoa.gr} 
\affiliation{Section of Solid State Physics, Department of Physics, School of Science, National
and Kapodistrian University of Athens, Panepistimiopolis, Zografos
157 84, Athens, Greece}
\affiliation{Solid Earth Physics Institute, Department of
Physics, School of Science, National and Kapodistrian University
of Athens, Panepistimiopolis, Zografos 157 84, Athens, Greece}

\author{Mary S. Lazaridou}
\affiliation{Solid Earth Physics Institute, Department of
Physics, School of Science, National and Kapodistrian University
of Athens, Panepistimiopolis, Zografos 157 84, Athens, Greece}

\begin{abstract}
By applying natural time analysis to the time
series of earthquakes, we find that the order parameter of
seismicity exhibits a unique change approximately at the date(s)
at which Seismic Electric Signals (SES) activities have been
reported to initiate. {In particular, we show that
the fluctuations of the order parameter of seismicity in Japan
exhibits a clearly detectable minimum approximately at the time of
the initiation of the SES activity observed 
almost two months before the onset of the Volcanic-seismic swarm
activity in 2000 in the Izu Island region, Japan.} To the best of
our knowledge, this is the first time that, well before the
occurrence of major earthquakes, anomalous changes are found to
appear almost simultaneously in two independent datasets of
different geophysical observables (geoelectrical measurements,
seismicity). {In addition, we show that these two
phenomena are also linked closely in space.}
\end{abstract}
\maketitle
\section{Introduction}
\label{sec1} Seismic
Electric Signals (SES) \citep{VAR84A,VAR96B} are low frequency ($\leq 1$Hz)
transient changes of the electric field of the Earth that precede
earthquakes. Several such  transient changes detected within a
short time are termed SES activity. A proper combination of the
SES physical properties enables the determination of the epicenter
and the magnitude of an impending earthquake (EQ)
\citep[]{VAR84A,VAR84B,VAR91,VAR93}. In addition,
 the small earthquakes subsequent to the initiation of an SES
 activity, when analyzed in a new time domain termed natural time (see below),
 enable the determination of the occurrence time of an impending
 mainshock a few days to around one week in advance
 \citep{NAT01,SAR08,SPRINGER}.

 Despite the successful predictions of several mainshocks in Greece, for
 example all the mainshocks with moment magnitude $M_w \geq$6.4
 during the decade 2001-2011
 \citep[see subsections 7.2.1 to  7.2.6 of][as well as a few major mainshocks in areas previously
 considered aseismic, see Chapters 5 and 14 of \citealt{MARY}]{SPRINGER}, the
 SES research has been a target of a heated debate, as noticed
 in a recent review by \citet{UYE09B}.
 It is the main objective of this paper to hopefully end
 this debate by reporting an important fact
 which unambiguously shows that the initiation of an SES activity  is accompanied
 by a clearly {detectable} change in an independent
 geophysical dataset of different physical observables.To understand the issue, we
 first summarize below the pressure stimulated
 polarization currents (PSPC) model proposed by \citet{VARBOOK} \citep[see also][]{VAR93} for
 the SES generation mechanism as well as  recapitulate the up to date knowledge
 on the lead time of the SES activities.

The PSPC model for the SES generation mechanism, based on Solid State Physics aspects,
 is consistent with the widely accepted
concept that the stress gradually increases in the future focal
region of an EQ. When this stress reaches a {\em critical} value,
a {\em cooperative} orientation of the electric dipoles (which are
anyhow present in the focal area due to  lattice defects in
the ionic constituents of the rocks, cf. all solids including metals, insulators and semiconductors contain intrinsic and extrinsic defects\cite{VAR74M,KOS75,VAR997,VAR80K133,VAR08438}) is attained, which leads to
the emission of a transient electric signal that constitutes an
SES (the {\em cooperativity} is a hallmark of criticality). Note
that no external electric field is a prerequisite for this
electric dipoles' orientation, because in the case of
inhomogeneous stress (which should occur during the EQ preparation
stage) the effect of the applied stress gradient is similar to that of an electric field \citep[see
also][]{VARNOV01}. The validity of this SES generation mechanism
is strengthened by the fact that the up to date experimental data
of SES activities (along with their associated magnetic field
variations) have been shown to exhibit infinitely ranged temporal
correlations \citep{NAT03A,NAT03B,NAT09V}, thus being in accord
with the conjecture of {\em critical} dynamics. According to
\citet{UYE09B}, the PSPC model is unique among other models in
that SES would be generated spontaneously during the gradual
increase of stress without requiring any sudden change of stress
such as microfracturing. The up to date observations of SES
activities in Japan \citep[for example see][]{UYE02,UYE09} as well
as in Mexico \citep[see][and references therein]{RAM11}  and in
California \citep[see][where magnetic field variations similar to
those associated with the SES activities in Greece have been
reported]{BER91,FRA90} have shown that their lead time is of the
order of a few months, in agreement with earlier observations in
Greece \citep{VAR91,VAR93,SPRINGER}. Thus, the  observations of
SES activities in various countries reveal that before the
occurrence of major earthquakes there is a {\em crucial} time
scale of around a few months or so \citep[up to around 5 months,
see][]{SPRINGER}, in which the {\em critical} stress is attained.
This may reflect that changes in the correlation properties of
other associated physical observables like seismicity may become
detectable at this time scale. It is exactly this aspect at which
our present work is focused on. In other words, it is the
objective of the present study to examine whether upon the
initiation of the emission of an SES activity there exists also a
noticeable change in the correlation properties of seismicity. To
unveil such a change we employ here natural time analysis (see
Section \ref{sec2}) since it has been demonstrated \citep[see][and
references therein]{SPRINGER} that novel dynamic features hidden
behind time series in complex systems emerge upon analyzing them
in natural time.

\section{Natural time analysis and seismicity. Background.} \label{sec2}
Natural time analysis, introduced a decade ago
\citep{NAT01,NAT02,NAT02A,NAT03B,NAT03A,NAT07}, has found applications in a large variety
of diverse fields and the relevant results have been compiled
in a recent monograph \citep{SPRINGER}.
In the case of seismicity, in a time series comprising $N$
earthquakes, the natural time $\chi_k = k/N$ serves as an index
for the occurrence of the $k$-th earthquake. The combination of this
index with the energy $Q_k$ released  during the $k$-th earthquake
of magnitude $M_k$, i.e., the pair $(\chi_k, Q_k)$, is studied in
natural time analysis. Alternatively, one studies the pair
$(\chi_k,p_k)$, where

\begin{equation}
p_k=\frac{Q_k}{\sum_{n=1}^NQ_n}
\end{equation}
stands for the normalized energy released during the $k$-th
earthquake. It has been found
\citep{NAT01,NAT03B,NAT03A,NAT05C,SPRINGER} that the variance of
$\chi$ weighted for $p_k$, designated by $\kappa_1$, which is
given by

\begin{equation}\label{kappa1}
\kappa_1=\sum_{k=1}^N p_k (\chi_k)^2- \left(\sum_{k=1}^N p_k
\chi_k \right)^2,
\end{equation}
plays a prominent role in natural time analysis. Note that $Q_k$,
and hence $p_k$, for earthquakes is estimated through the usual
relation \citep{KAN78}

\begin{equation}\label{qkprop}
Q_k \propto 10^{1.5M_k}
\end{equation}

Seismicity exhibits complex correlations in time, space and
magnitude ($M$) that have been studied in numerous investigations
\citep[for example, see][]{BAK02,LEN11,LIP12,SARCHRIS12A}. The
observed earthquake scaling laws \citep[e.g. see][]{TUR97} are
widely accepted to indicate the existence of phenomena closely
associated with the proximity of the system to a critical point
\citep{HOL06,SOR00,CAR94,XIA08}. Here, we take the view that
earthquakes are (non-equilibrium) critical phenomena, and employ
the analysis in natural time $\chi$, because in the frame of this
analysis an order parameter for seismicity has been introduced. In
particular, it has been explained \citep{NAT05C}  in detail
\citep[see also pp. 249-253 of][]{SPRINGER} that the quantity
$\kappa_1$ given by Eq. (\ref{kappa1}) -or the normalized power
spectrum in natural time $\Pi(\omega)$ as defined by
\citet{NAT01,NAT02} for natural angular frequency $\omega
\rightarrow 0$- can be considered as an order parameter for
seismicity since its value changes abruptly when a mainshock (the
new phase) occurs, and in addition the statistical properties of
its fluctuations resemble those in other nonequilibrium and
equilibrium critical systems.

The value of the order parameter ($\kappa_1$) itself plays a key
role in identifying the occurrence time of a mainshock. This is
so, because it has been found \citep{NAT01,SAR08,SPRINGER} that a
mainshock occurs in a few days to one week after the $\kappa_1$
value is recognized to have approached 0.070 in the natural time
analysis of the seismicity subsequent to the initiation of an SES
activity. This has been ascertained in several major mainshocks in
various countries including Greece, Japan and USA \citep[see][and
references therein]{SPRINGER}.

\section{The data analyzed and the procedure followed.} \label{sec4}

To achieve the goal of the present study we need two types of
datasets. The one should be a clearly observed SES activity and
the other  an authoritative earthquake catalog that will include
the time series of the earthquakes during the time period of the
observation of this SES activity. In Greece, there are several
pairs of such datasets since our continuous SES observations are
lasting for almost 30 years. However, in order to make our
presentation more objective, we intentionally consider datasets
reported by other independent workers that are easily accessible
from the international literature. Specifically we shall consider
the well known SES activity published by Uyeda and coworkers
\citep{UYE02,UYE09} that preceded the volcanic-seismic swarm
activity in 2000 in the Izu Island region, Japan. This was a
pronounced SES activity with innumerable signals that started
almost two months prior to the swarm onset. (The precise date of
its initiation was reported to be on 26 April 2000, but see also
below.) This swarm was then characterized by Japan Meteorological
Agency (JMA) as being the largest earthquake swarm ever recorded
\citep{JMA00}. Further, to meet our goal we analyzed in natural
time the series of the earthquakes reported during this period by
the JMA seismic catalog. In particular, we considered all EQs
within the area $25^o-46^o$N, $125^o-146^o$E, which covers the
whole Japanese region \citep[for example see][]{TAN04}. The
seismic moment $M_0$, which is proportional to the energy released
during an EQ (this is the quantity $Q_k$ of the $k$-th event used
in natural time analysis), was obtained from the magnitude $M_{\rm
JMA}$ reported in the JMA catalog by using the approximate
formulae of \citet{TAN04} that interconnect $M_{\rm JMA}$ with
$M_w$. Setting a threshold $M_{\rm JMA} > 3.4$ to assure data
completeness, there exist 52,718 EQs in the period from 1967 to
the time of Tohoku EQ. This reflects that we have on the average
$\sim 10^2$ EQs per month.

Concerning the procedure followed, we consider a sliding natural
time window of fixed length comprising $W$ consecutive events.
Starting from the first earthquake, we calculate the $\kappa_1$
values using $N =$ 6 to 40 consecutive events (cf. This natural time window is used here as well as in \cite{TECTO12}, while in \cite{PNAS13,PNAS15} $N$ reaches values up to $W$). We next turn to the
second earthquake, and repeat the calculation of $\kappa_1$. After
sliding, event by event, through the whole natural time window,
the computed $\kappa_1$ values enable the calculation of their
average value $\mu(\kappa_1)$ and their standard deviation
$\sigma(\kappa_1)$ that correspond to this natural time window of
length $W$. We then determine the variability $\beta$ of
$\kappa_1$, i.e., the quantity $\beta$ defined \citep{NEWEPL} as

\begin{equation}\beta =\frac{ \sigma(\kappa_1)}{\mu(\kappa_1)}.
\end{equation}

In order to simplify the discussion of our results we employ the
following change \citep{VAR11}: For each earthquake $e_i$ in the
seismic catalog, we calculate the $\kappa_1$ values resulting when
using the {\em previous} 6 to 40 consecutive earthquakes. Then,
the hitherto obtained $\kappa_1$ values for the earthquakes
$e_{i-W+1}$ to $e_i$ were considered for the estimation of the
variability $\beta$ for a natural time window length $W$. The
resulting $\beta$ value, {labelled} $\beta_i$, was attributed to
$e_i$, the data of which was obviously {\em not included} in the
$\beta_i$ estimation.

Since we are interested -as explained in the first Section- on
time scales comparable to that of the lead time of an SES
activity, after considering that in Japan  we have $\sim 10^2$ EQs
with $M > 3.4$ per month, as mentioned, we employ here the
following natural time window lengths: $W=$ 100, 200, 300, and 400
earthquakes.

\section{Results }\label{sec5}
We first present the results of our analysis during nine months
{(see also the Appendix)} until just before the occurrence of the
$M$6.5 EQ on 1 July 2000 close to Niijima Island.  During this
period, i.e., 1 October 1999 to 1 July 2000, the four curves in
Fig.\ref{fig1}(a) depict the computed values of the variability
$\beta$ of the order parameter $\kappa_1$ for seismicity versus
the conventional time (LT) for $W$=100 (red), 200 (green), 300
(blue) and 400 (magenta). We see that the variability $\beta$
remains more or less constant until around 20 March 2000 and
thereafter starts to decrease \citep[note that][in their Fig. 11,
have noticed that SES but of very small amplitude initiated
approximately on 22 March]{UYE09}. A clear minimum in the $\beta$
variability of $\kappa_1$ is subsequently observed around the last
week of April. To better visualize it, we now depict in expanded
time scale an excerpt of Fig.\ref{fig1}(a), see Fig.\ref{fig1}(b),
that refers only to the period from 1 February 2000 to 1 July
2000, i.e., during five months until just before the occurrence of
the M6.5 EQ on 1 July 2000. An inspection of Fig.\ref{fig1}(b)
reveals that the minimum of the variability $\beta$ of $\kappa_1$
is observed on the following date(s): 23 April for $W=100$, 26
April for $W=200$, 21-23 April for $W=300$, and 23-24 April for
$W=400$ (these dates are marked on the figure). For $W=100$, 300,
and 400, it seems that there is a tendency for the date of the
minimum of $\beta$ to precede somewhat that of the initiation of
the SES activity by a few days, but for $W=200$ the two dates
coincide. In view of the experimental errors, however, mainly due
to the earthquake magnitude determination, we cannot make an
assertion on an exact coincidence of the two dates. In other
words, under the current experimental uncertainty (of around a few
days or so), we can say that our main finding points to the fact
that the fluctuations of the order parameter $\kappa_1$ of
seismicity became minimum around (or at least very close to) the
date (26 April) of the initiation of the  SES activity reported by
\citet{UYE09}. A simple calculation shows that the probability of
ascribing this almost simultaneous appearance of the two phenomena
to chance, is very small  if we just take into account that
Fig.\ref{fig1}(a) extends over a nine-month period (i.e., of the
order of 1\% even when considering a single value of $W$).

Note also that in Fig.\ref{fig1}(b), after the beginning of June
an increase of the variability $\beta$ becomes evident, but on 27
June 2000  (approximately at 11:44 LT) an abrupt decrease occurs.
This may be understood in the following context: By setting
natural time zero at the initiation time of the SES activity,
\citet{UYE09} conducted the natural time analysis of seismic
events in the rectangular region from $N 33.7^o$ to $N 34.8^o$ and
from $E 139^o$ to $E140^o$ depicted in the inset of Fig.
\ref{fig2}. They found that on 27 June, the order parameter
$\kappa_1$ approached {\em gradually}, i.e., without significant
fluctuations, see their Fig. 7(b) \citep[in a similar fashion as
observed repeatedly in the Greek cases, see][]{NAT01} the value
0.070, thus signalling the imminent mainshock on 1 July 2000.

\section{{The robustness of the co-location in time of the two phenomena with respect to the area selection}}
{The aforementioned results were obtained as
mentioned by analyzing in natural time the seismicity within the
wide area 25-46$^o$N 125-146$^o$E,i.e., $21^o \times 21^o$. In
this section, we shall focus on studying how robust is the
co-location in time of the two phenomena with respect to the
choice of area selection. In other words, we shall present below
an analysis that investigates how the date of the
minimum of $\beta$ varies if the area selection is changed. Along these
lines, we employ a sliding area window and determine the
occurrences of the minimum of $\beta$ as a function of date. The
analysis has been carried out for different sizes of the sliding
area window.}

{Before showing the present results of our
analysis, we mention two earlier results. First, the wide area
$N_{25}^{46} E_{125}^{146}$ used here (surrounded by the yellow
square in Fig.\ref{newfig2}) has been already employed by
\citet{NAT05C,NAT06B} in order to show that the statistical
properties of the fluctuations of the order parameter $\kappa_1$
of seismicity resemble those in other nonequilibrium and
equilibrium critical systems as already mentioned above in Section
\ref{sec2}. In particular, the properties of the probability
density function (pdf) P$(\kappa_1)$ versus $\kappa_1$ -obtained
by means of the procedure described in Section \ref{sec4}- for the
{\em long term} seismicity in the area $N_{25}^{46} E_{125}^{146}$
were studied. It was found \citep{NAT05C} that the {\em scaled}
distribution P$(y)\equiv \sigma(\kappa_1)$P$(\kappa_1)$ plotted
versus $y \equiv (\mu(\kappa_1)-\kappa_1)/ \sigma(\kappa_1)$ of
this area falls on the {\em same} curve ({\em universal}) with the
ones obtained from different seismic areas upon using the
corresponding earthquake catalogs, e.g., Southern California (as
well as that of the worldwide seismicity). This ``universal''
curve for the long term seismicity exhibits strikingly similar
features (for example a common ``exponential tail'' characteristic
of rare non-Gaussian fluctuations, e.g., of greater than six
standard deviations from the mean) with the order parameter
fluctuations in other nonequilibrium systems (e.g., 3D turbulent
flow) as well as in several equilibrium critical systems (e.g., 2D
Ising, 3D Ising). Second, we mention a study just published the
findings of which will be of usefulness in an attempt towards
understanding the results of our analysis that will be presented
below. Specifically, \citet*{TEN12} proposed and developed a
network approach to earthquake events. In this network, a node
represents a spatial location while a link between two nodes
represents similar activity patterns in the two different
locations. The strength of a link is proportional to the strength
of the cross-correlation in activities of two nodes joined by the
link. They applied this network approach to the Japanese
earthquake activity spanning the 14 year period 1985-1998 within
an area $23^o \times 23^o$ that exceeds slightly (i.e., by $2^o$
in latitude and $2^o$ in longitude) the area used by our group.
\citet{TEN12} found strong links representing large correlations
between patterns in locations separated by {\em more than 1000}
km. They found network characteristics not attributable to chance
alone, including a large number of network links, high node
assortativity, and strong stability over time. The network links
(along with the corresponding nodes) identified by \citet{TEN12},
see their Fig.6(a), are superimposed on a map of the Japanese
archipelago in Fig.\ref{newfig2}.}

{In the map of Fig.\ref{newfig2}, we also mark with
green stars the epicenters of the 200 events when the ending of
the natural time window of length $W=$200 lies on the minimum of
the $\kappa_1$ variability marked in the green curve in
Fig.\ref{fig1}(b), i.e., on 26 April 2000. Let us now examine what
happens with this date when our analysis is carried out for
different sizes of the sliding area window (we follow, of course,
the same procedure as in the wide area for $W$=200, i.e.,  for
each size we consider as $W$ the corresponding number of the
events that would on average occur  within two months).  At the
moment, we restrict ourselves to sizes which correspond to
distances that markedly exceed the aforementioned distance of 1000
km. In particular, we present below examples for sliding area
window sizes $20^o \times 20^o$, $17.5^o \times 17.5^o$ and $15^o
\times 15^o$. The results obtained for the period 1 October 1999
to 1 July, 2000 (i.e., in a similar fashion as in
Fig.\ref{fig1}(a)) are as follows:}

{For the sliding area window $20^o \times 20^o$,
the resulting curves for the variability $\beta$ versus the
conventional time for the four areas $N_{25}^{45} E_{125}^{145}$,
$N_{26}^{46} E_{125}^{145}$, $N_{25}^{45} E_{126}^{146}$, and
$N_{26}^{46} E_{126}^{146}$ are shown in Fig.\ref{newfig3} in red,
green, blue and cyan, respectively. All these four curves exhibit a
minimum on 26 April 2000, thus agreeing with the findings in the
green curve in Fig.\ref{fig1} for $W=$200 for the area
$N_{25}^{46} E_{125}^{146}$. The curve for the latter case is also
shown (thick black) in Fig.\ref{newfig3} for the sake of
comparison.}

{For the sliding area window $17.5^o \times 17.5^o$
the four areas investigated are the following:  $N_{25.0}^{42.5}
E_{125.0}^{142.5}$, $N_{27.0}^{44.5} E_{125.0}^{142.5}$,
$N_{25.0}^{42.5} E_{127.0}^{144.5}$ and $N_{27.0}^{44.5}
E_{127.0}^{144.5}$ leading to the corresponding curves plotted in
Fig.\ref{newfig4} in red, green, blue and cyan, respectively. They
show that the minimum of $\beta$ is identified on the following
dates: 30 April, 21 April, 21 April and 26 April 2000,
respectively, which are more or less in agreement with the date(s)
in Fig.\ref{fig1}. By the same token as in Fig.\ref{newfig3}, we
insert a thick black curve in Fig.\ref{newfig4} showing the
results plotted in Fig.\ref{fig1} for $W=$200.}

{As a third example, we present the results for the
sliding area window $15^o \times 15^o$. In this case, the study
included the investigation of sixteen areas -separated by $2^o$ in
longitude and/or $2^o$ in latitude- the results of which are
depicted in Figs.\ref{newfig5}(a), (b), (c) and (d). The
coordinates of each of these areas are shown in  the upper left
corner in each figure, which also includes  a thick black curve
showing the result of the original area $N_{25}^{46}
E_{125}^{146}$ for $W=$200 for the sake of comparison. The results
could be summarized as follows: In ten areas (out of 16), we find
that the minimum of $\beta$ appears at dates lying between 21 and
27 April 2000 (i.e., see the three cases marked in each of the
figures \ref{newfig5}(a), (b) and (c) as well as the one case
marked in red in Fig.\ref{newfig5}(d)) and hence close to the date
of 26 April 2000 observed in Fig.\ref{fig1} for $W=$200). In four
areas (out of 16) the minimum of $\beta$ appears somewhat shifted,
i.e., around 8 May 2000 (these are the three cases marked with a
brown box in Fig.\ref{newfig5}(d) and the green curve in
Fig.\ref{newfig5}(c)). Two areas, however, i.e., $N_{31}^{46}
E_{125}^{140}$ and $N_{29}^{44} E_{125}^{140}$, the results of
which are depicted in red in Figs.\ref{newfig5}(a) and (b), do not
exhibit a minimum of $\beta$ close to the expected date of 26
April 2000. In an attempt to understand why these two areas behave
differently compared to the other areas, they are shown in the map
of Fig.\ref{newfig2} with broken black and solid purple squares,
respectively superimposed on the network links map identified by
\citet{TEN12}. For the sake of comparison, we also show in the
same figure, as an example, the area $N_{31}^{46} E_{131}^{146}$
-see the rightmost green square in Fig.\ref{newfig2}- in which our
analysis led to the conclusion that there exists a minimum in the
variability $\beta$ on 26 April 2000, see the cyan curve in
Fig.\ref{newfig5}(a), thus being in agreement with the date of the
minimum (26 April) identified in Fig.\ref{fig1} (green curve) for
$W=$200. A careful inspection of Fig.\ref{newfig2} reveals that a
number of nodes - three of which are shown with arrows- associated
with a multitude of links lie outside of the two areas
$N_{31}^{46} E_{125}^{140}$ and $N_{29}^{44} E_{125}^{140}$ but
inside the area $N_{31}^{46} E_{131}^{146}$.}

\section{{Investigation on whether the two phenomena are linked also in space}}\label{newsec6}
{To answer this important question, we must focus
on an analysis similar to that presented in the previous Section,
but to a sliding area window of appreciably smaller size. This
size cannot be smaller than $5^o \times 5^o$ if we consider the
following: In the whole area $21^o \times 21^o$ we have 200 events
-covering almost two months- that preceded the minimum on 26 April
2000 shown in the green curve of Fig.\ref{fig1}, hence an area
window of $5^o \times 5^o$ corresponds on the average to
$\frac{200}{(21^o \times 21^o)/(5^o \times 5^o)}\approx 11$ events
per two months.  In addition, we have to take into account that in
order to identify the date of the occurrence of the minimum of
$\beta$ in an area window of size $5^o \times 5^o$ with an
uncertainty of around a few days, we must have at least, on the
average, 2 events per week; thus, in an area window $5^o \times
5^o$ we must have at least 16 events for an almost two months
period (8 weeks).  }

{An area window of size $5^o \times 5^o$ sliding
through the area $N_{25}^{46} E_{125}^{146}$  results in $9\times
9$ areas when varying the center of each area with steps of $2^o$
in  longitude and/or $2^o$ in latitude. Among these 81 areas, we select
those that have at least 16 events per two months and find the 23
areas the coordinates of which are given in Table \ref{tab1} and
depicted in Fig.\ref{newfig6}(a). By analyzing these 23 areas we
find the following:}

{First,  there exist only three areas labeled
``65'', ``87'' and ``97'', i.e., the ones with coordinates
$N_{33}^{38} E_{135}^{140}$, $N_{37}^{42} E_{139}^{144}$ and
$N_{37}^{42} E_{141}^{146}$, that result in a clearly observable
minimum of the variability $\beta$ the date of which differs by no
more than a few days from the date 26 April 2000 exhibited in
Fig.\ref{fig1} (green curve) for $W=200$. In particular, the
minimum in these three areas is observed on 25, 26, and 26 April,
respectively, as shown in Fig.\ref{newfig6}(b) see the curves
plotted in blue, green and cyan. Obviously, the lowest minimum is
exhibited by the first area, i.e., $N_{33}^{38} E_{135}^{140}$,
which remarkably includes the epicenters -marked with asterisks in
Fig.\ref{newfig6}(c)- of the two earthquakes that occurred on 1
and 30 July 2000 close to Niijima Island measuring station (see
also the inset of Fig.\ref{fig2}).}

{Second,  concerning the area labelled ``64'', i.e.,
$N_{31}^{36} E_{135}^{140}$, which (also includes, see Fig.\ref{newfig6}(c),
the epicenters of the aforementioned two EQs and) is somewhat
displaced to the south compared to the area $N_{33}^{38}
E_{135}^{140}$ mentioned above, we observe the following: It
exhibits a very shallow minimum of the variability $\beta$ around
15 April 2000, which however is almost a plateau extending up to 24 April 2000, hence being more or less close to the date (26
April) of the minimum observed in Fig.\ref{fig1} (green curve) for
$W=$200.}

{In other words, the findings of this Section,
could  be summarized as follows: Here, by using a narrow $5^o
\times 5^o$ spatial window sliding through the wide
($21^o\times21^o$) area  $N_{25}^{46} E_{125}^{146}$,  we
investigated the earthquake events that would occur in an almost
two months period. Recall that in the wide area these earthquakes
have been interpreted (in the discussion of the minimum in the
green curve of Fig.\ref{fig1}) to correlate with the SES.
 We found that the characteristics of the
fluctuations of  $\kappa_1$ (i.e., their lowest
minimum) are linked to the seismicity occurring in the areas
$N_{33}^{38} E_{135}^{140}$ and $N_{31}^{36} E_{135}^{140}$ that
include the Izu Island region which became active
a few months later.}

\section{{Conclusions}}\label{sec6}
Just by analyzing the Japanese seismic catalog in
natural time, and employing a sliding natural time window
comprising the number of events that would occur in a few months,
we find the following: The fluctuations of the order parameter
$\kappa_1$ of seismicity exhibit a clearly detectable minimum
approximately at the time of the initiation of the pronounced SES
activity  observed by \citet{UYE02,UYE09} almost two months before
the onset of the volcanic-seismic swarm activity in 2000 in the
Izu Island region, Japan. This reflects that presumably the same
physical cause led to both effects observed, i.e, the emission of
the SES activity and the change of the correlation properties
between the earthquakes. This might be the case  when the stress
reached its {\em critical} value, if we think in terms of the SES
generation model proposed by \citet{VARBOOK}. {In
addition, the two phenomena discussed are found to be also linked
in space.}

Finally, we note that the appearance of minima in the variability
$\beta$ of $\kappa_1$ before major earthquakes in Japan is
investigated in detail elsewhere \citep{PNAS13}. For the vast
majority of these cases, however, the main conclusion of the
present investigation, i.e., the almost simultaneous appearance of
these minima with the initiation of SES activities, cannot be
checked  due to the lack of geolectrical data. It is this lack of
data which obliges us, as explained in the Appendix, to present in
Fig. \ref{fig1}(a) the results of our analysis solely for a period
of nine months, i.e., 1 October 1999 to 1 July 2000.

\begin{figure*}
\includegraphics[scale=0.7]{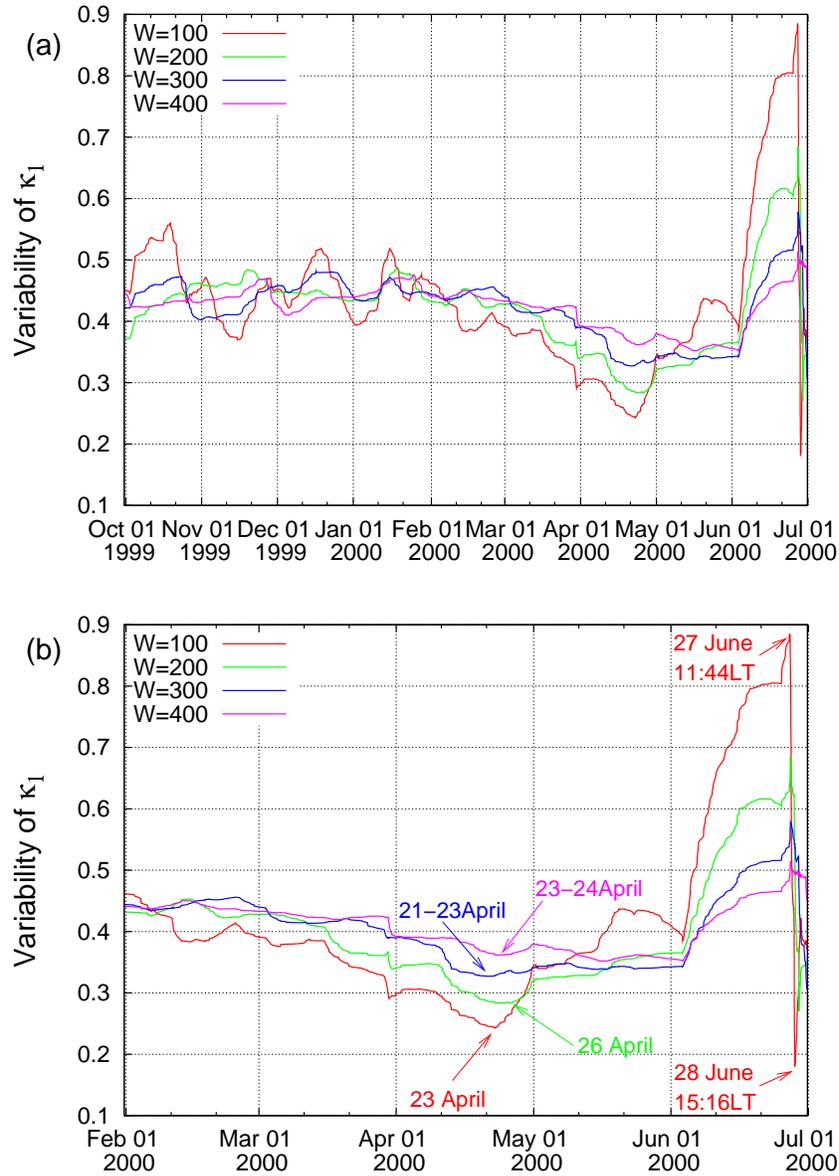}
\caption{(color online) The variability $\beta$ of the order
parameter $\kappa_1$ of seismicity versus the conventional time
(UT) for natural time window lengths $W$=100 events (red), $W$=200
events (green), $W$=300 events (blue) and $W$=400 events (magenta)
during the period from (a):1 October 1999 to 1 July 2000 and (b):1
February 2000 to 1 July 2000 until just before the occurrence of
the M6.5 EQ on 1 July 2000. The JMA catalog was used with
magnitude threshold $M > 3.4$}\label{fig1}
\end{figure*}

\begin{figure*}
\includegraphics[scale=0.7]{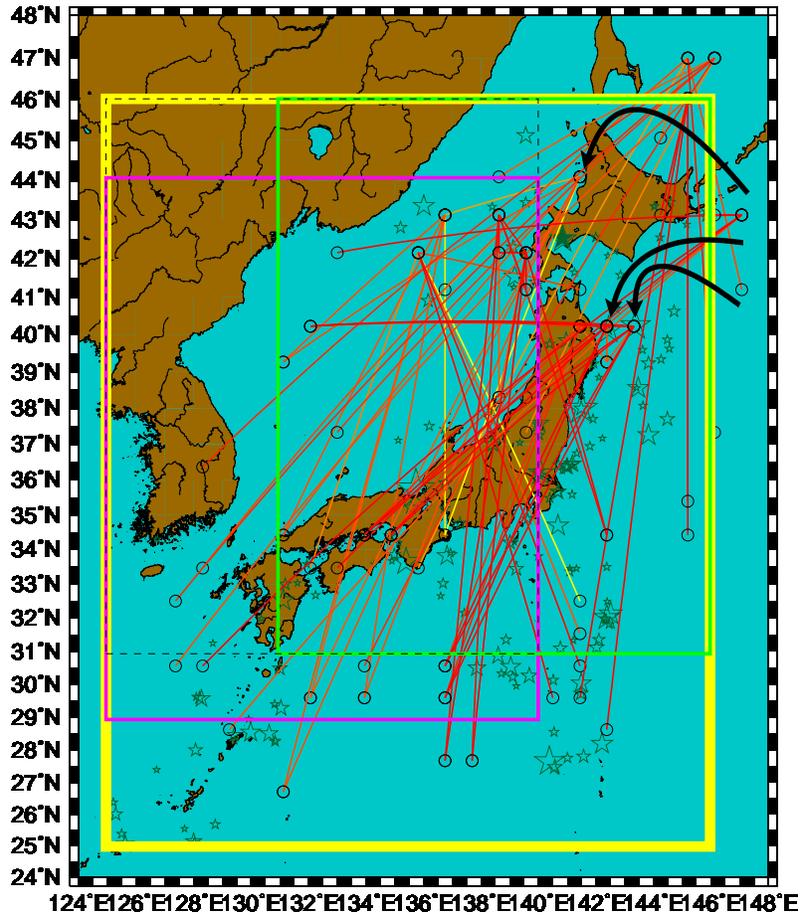}
\caption{(color online) {Network links, as reported
by \citet{TEN12}, superimposed on a map of the Japanese
archipelago. The asterisks show the epicenters of the 200 events
when the ending of the natural time window of length $W=$200 lies
on the minimum of $\beta$ on 26 April 2000, see the green curve in
Fig.\ref{fig1}(b). The yellow square boundaries mark the area
$N_{25}^{46} E_{125}^{146}$ used in the present analysis, to
obtain the results in Fig.\ref{fig1}. The two areas (out of 16),
$N_{31}^{46} E_{125}^{140}$ (broken black) and $N_{29}^{44}
E_{125}^{140}$ (solid purple), which when using a sliding area
window $15^o \times 15^o$ do not exhibit a minimum of $\beta$ at a
date close to 26 April 2000, while the area $N_{31}^{46}
E_{131}^{146}$ (solid green) does so.}}\label{newfig2}
\end{figure*}

\begin{figure*}
\includegraphics[scale=0.5]{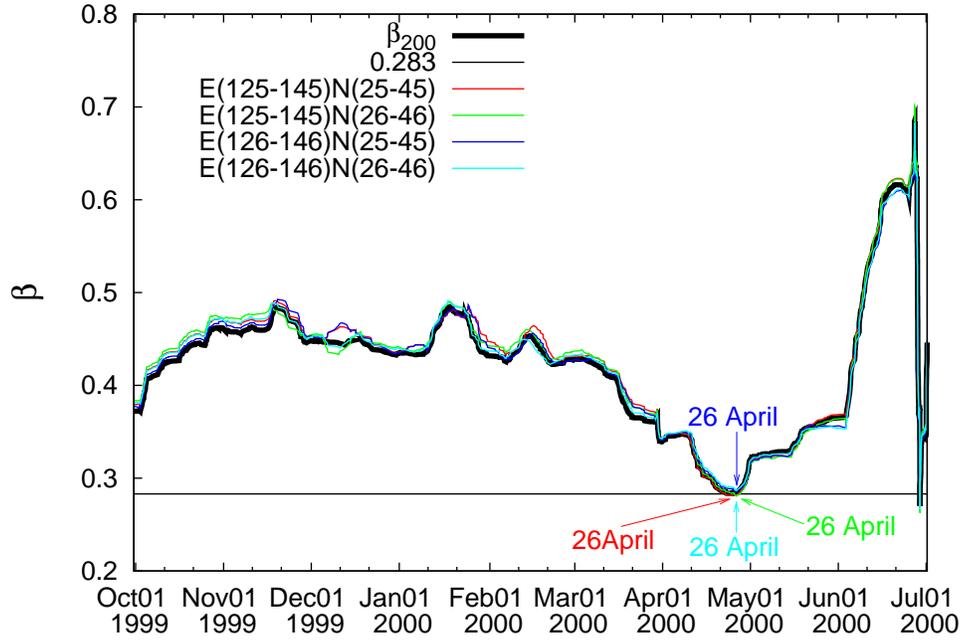}
\caption{(color online) {Results of the analysis
leading to the variability $\beta$ versus the conventional time
when using a sliding area window $20^o \times 20^o$. The curves
shown in red, green, blue and cyan depict the corresponding
results for the areas $N_{25}^{45} E_{125}^{145}$, $N_{26}^{46}
E_{125}^{145}$, $N_{25}^{45} E_{126}^{146}$ and $N_{26}^{46}
E_{126}^{146}$. For the sake of comparison, the thick black curve
reproduces the results of Fig.\ref{fig1} for a natural time window
of length $W=$200 sliding through the whole area $N_{25}^{46}
E_{125}^{146}$.}}\label{newfig3}
\end{figure*}

\begin{figure*}
\includegraphics[scale=0.5]{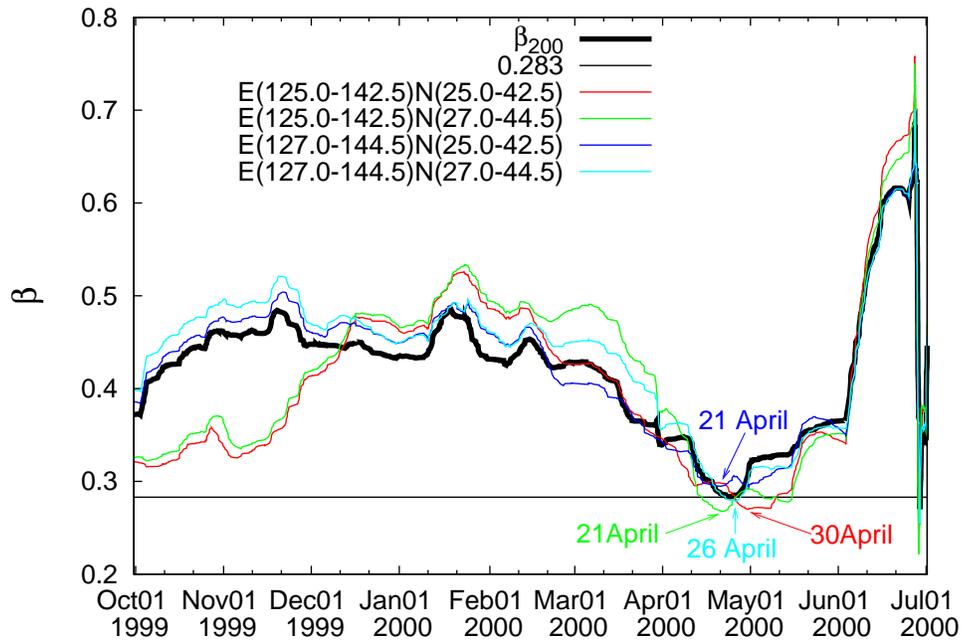}
\caption{(color online) {The same as
Fig.\ref{newfig3}, but for a sliding area window $17.5^o \times
17.5^o$. The results shown correspond to the following areas:
$N_{25.0}^{42.5} E_{125.0}^{142.5}$ (red), $N_{27.0}^{44.5}
E_{125.0}^{142.5}$ (green), $N_{25.0}^{42.5} E_{127.0}^{144.5}$
(blue) and $N_{27.0}^{44.5} E_{127.0}^{144.5}$
(cyan).}}\label{newfig4}
\end{figure*}

\includegraphics[scale=0.8]{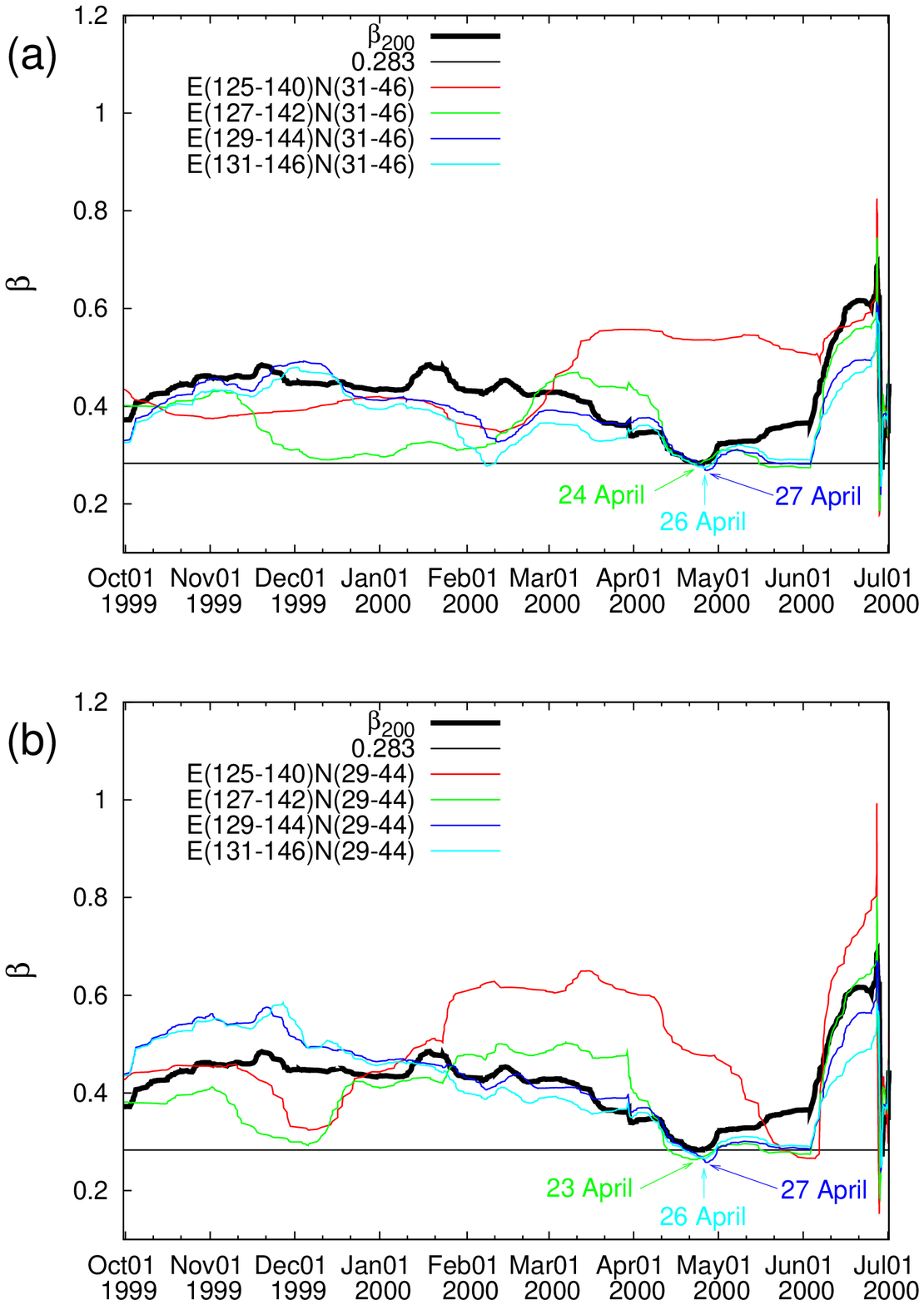}

\begin{figure*}
\includegraphics[scale=0.8]{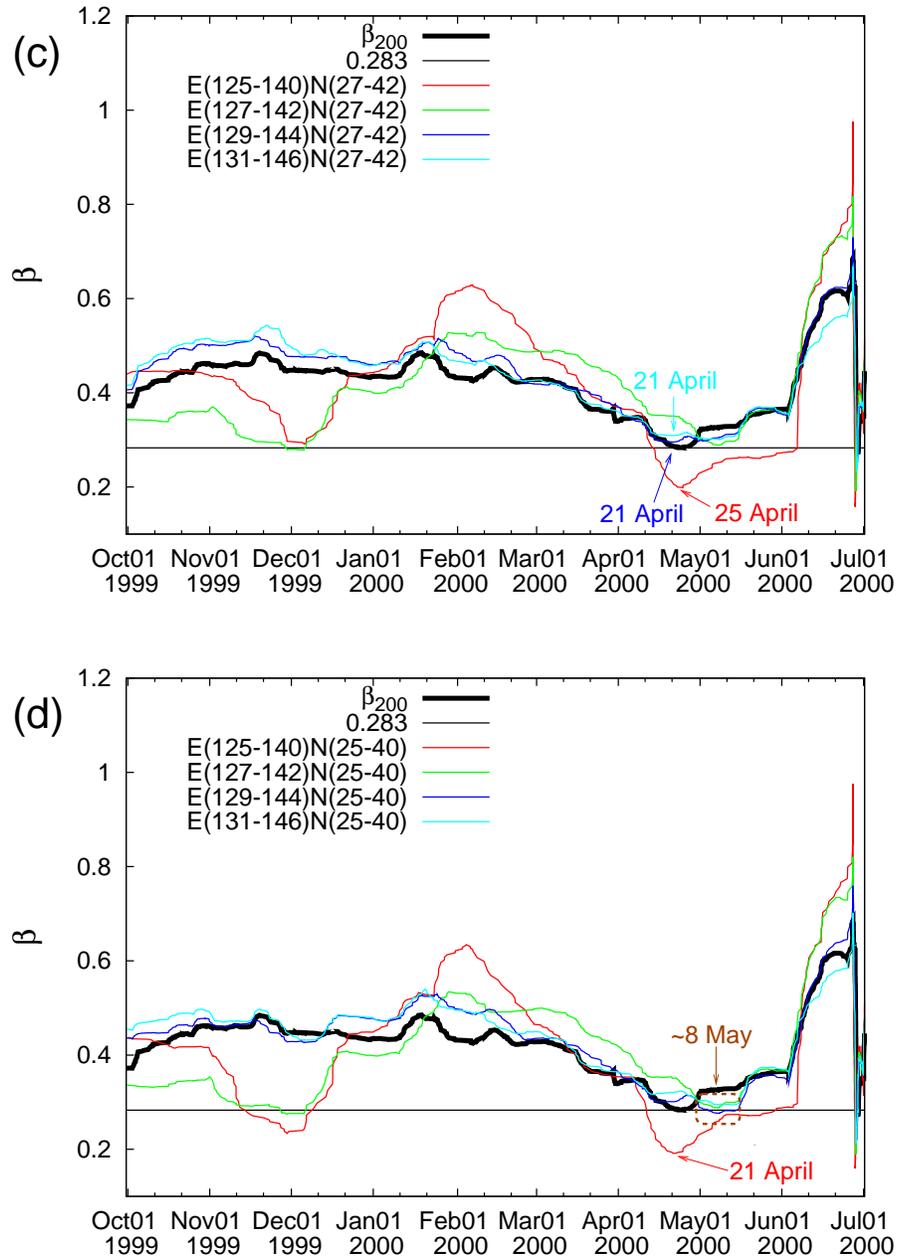}
\caption{(color online) {The same as
Fig.\ref{newfig3}, but for a sliding area window $15^o \times
15^o$. The corresponding areas for which the results are depicted
in (a), (b), (c) and (d) are written in the uppermost left corner
in each figure).}}\label{newfig5}
\end{figure*}

\includegraphics[scale=0.8]{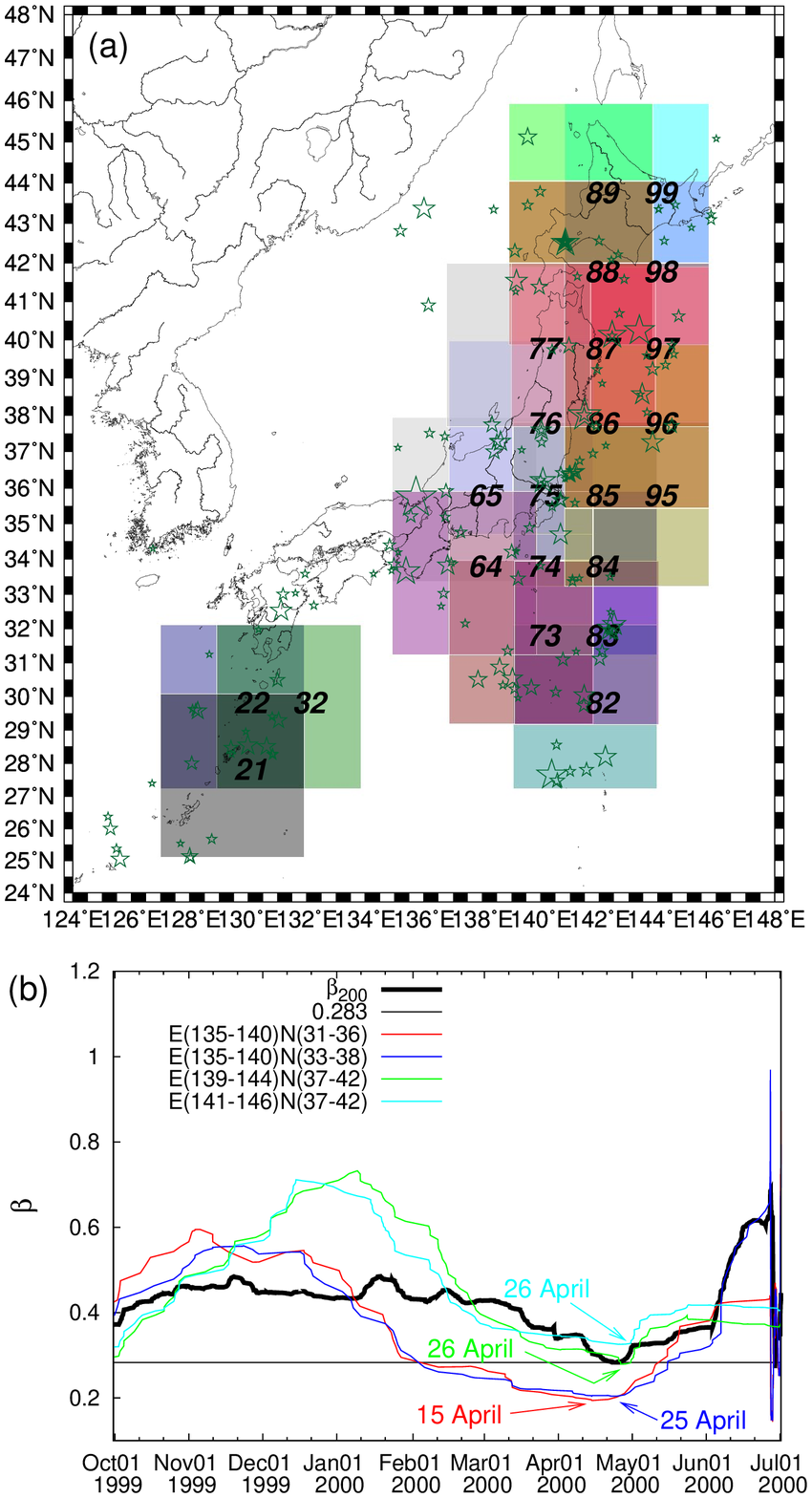}
\begin{figure*}
\includegraphics[scale=0.8]{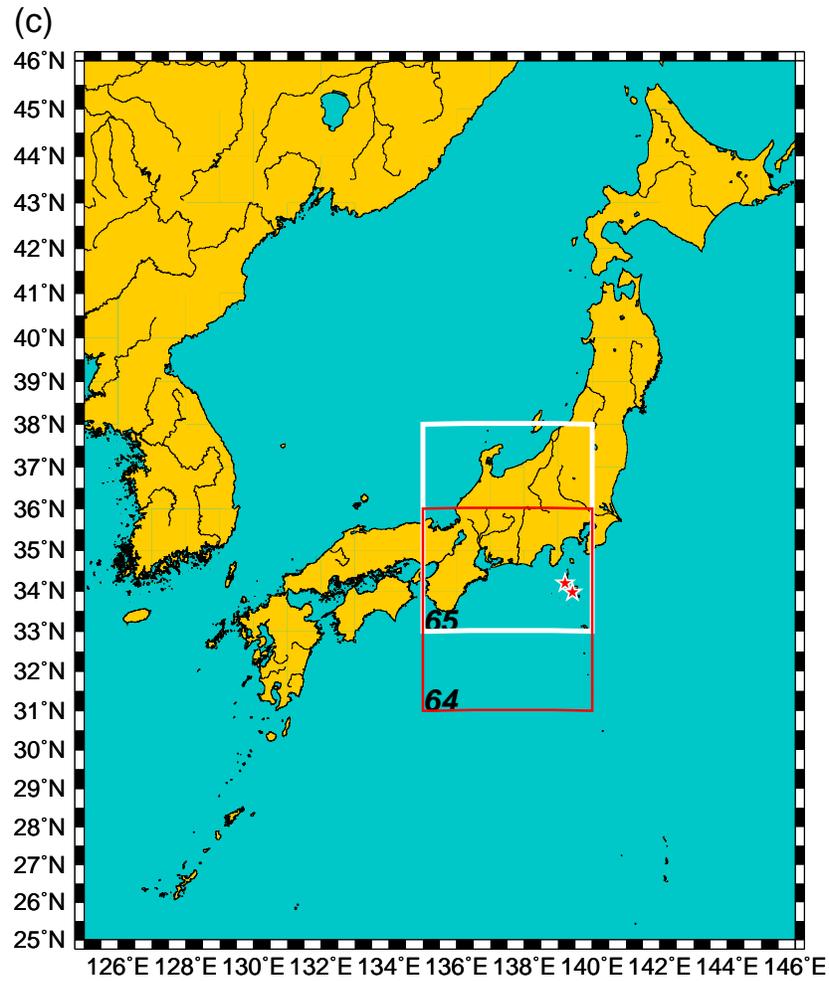}
\caption{(color online) {Results of the analysis
when using a sliding area window $5^o \times 5^o$. (a)Map showing
the areas in each of which at least 16 events occur during an
almost two months period (see Table \ref{tab1}). (b)The
variability $\beta$ versus the conventional time in those areas
that exhibited a minimum close to the date 26 April 2000.  (c)Map
showing the two  areas labeled ``65'' and ``64'', i.e,
$N_{33}^{38} E_{135}^{140}$ (white) and  $N_{31}^{36}
E_{135}^{140}$ (red), that among the four areas in (b) exhibit the
smaller variability and include the epicenters (marked with stars)
of the two earthquakes that occurred on 1 and 30 July 2000 (see
the text).}}\label{newfig6}
\end{figure*}

\begin{table}
\caption{{Investigation by means of a sliding area
window $5^o \times 5^o$. List of the 23 areas, see also
Fig.\ref{newfig6}(a), in each of which we have on the average at
least 16 events in an almost two months period.}} \label{tab1}
\begin{tabular}{cccccc}
\hline Area label & LONG ($^o$E) & LAT ($^o$N) & Center of the area &  Average Nr. of events \\
&&&&per 2 months \\
\hline
 21 & 127-132& 25-30& 129.5E   27.5N &   17 \\
 22 & 127-132& 27-32& 129.5E   29.5N &   18 \\
 32 & 129-134& 27-32& 131.5E   29.5N &   16 \\
 64 & 135-140& 31-36& 137.5E   33.5N &   29 \\
 65 & 135-140& 33-38& 137.5E   35.5N &   33 \\
 73 & 137-142& 29-34& 139.5E   31.5N &   22 \\
 74 & 137-142& 31-36& 139.5E   33.5N &   40 \\
 75 & 137-142& 33-38& 139.5E   35.5N &   53 \\
 76 & 137-142& 35-40& 139.5E   37.5N &   35 \\
 77 & 137-142& 37-42& 139.5E   39.5N &   22 \\
 82 & 139-144& 27-32& 141.5E   29.5N &   20 \\
 83 & 139-144& 29-34& 141.5E   31.5N &   23 \\
 84 & 139-144& 31-36& 141.5E   33.5N &   38 \\
 85 & 139-144& 33-38& 141.5E   35.5N &   52 \\
 86 & 139-144& 35-40& 141.5E   37.5N &   48 \\
 87 & 139-144& 37-42& 141.5E   39.5N &   45 \\
 88 & 139-144& 39-44& 141.5E   41.5N &   37 \\
 89 & 139-144& 41-46& 141.5E   43.5N &   19 \\
 95 & 141-146& 33-38& 143.5E   35.5N &   25 \\
 96 & 141-146& 35-40& 143.5E   37.5N &   38 \\
 97 & 141-146& 37-42& 143.5E   39.5N &   44 \\
 98 & 141-146& 39-44& 143.5E   41.5N &   39 \\
 99 & 141-146& 41-46& 143.5E   43.5N &   22 \\
\hline
\end{tabular}
\end{table}

\clearpage

\appendix
\section{{Additional comments on points discussed in the text.}}

An analysis similar to that shown in Fig. \ref{fig1}(a), is
presented by \citet{PNAS13} for an appreciably longer period,
i.e., from 1 January 1984 until the Tohoku earthquake on 11 March
2011. This is so, because the interconnection between the minima
of the variability $\beta$ of the order parameter of seismicity
and major earthquakes investigated by \citet{PNAS13} requires
solely the knowledge of seismic data. On the other hand, to meet
the purpose of the present study we need {as
mentioned} {\em both} seismic and geoelectrical data. The lack of
the latter data imposes certain constraints on our study that will
become clear below.

\begin{figure}
\includegraphics[scale=0.7]{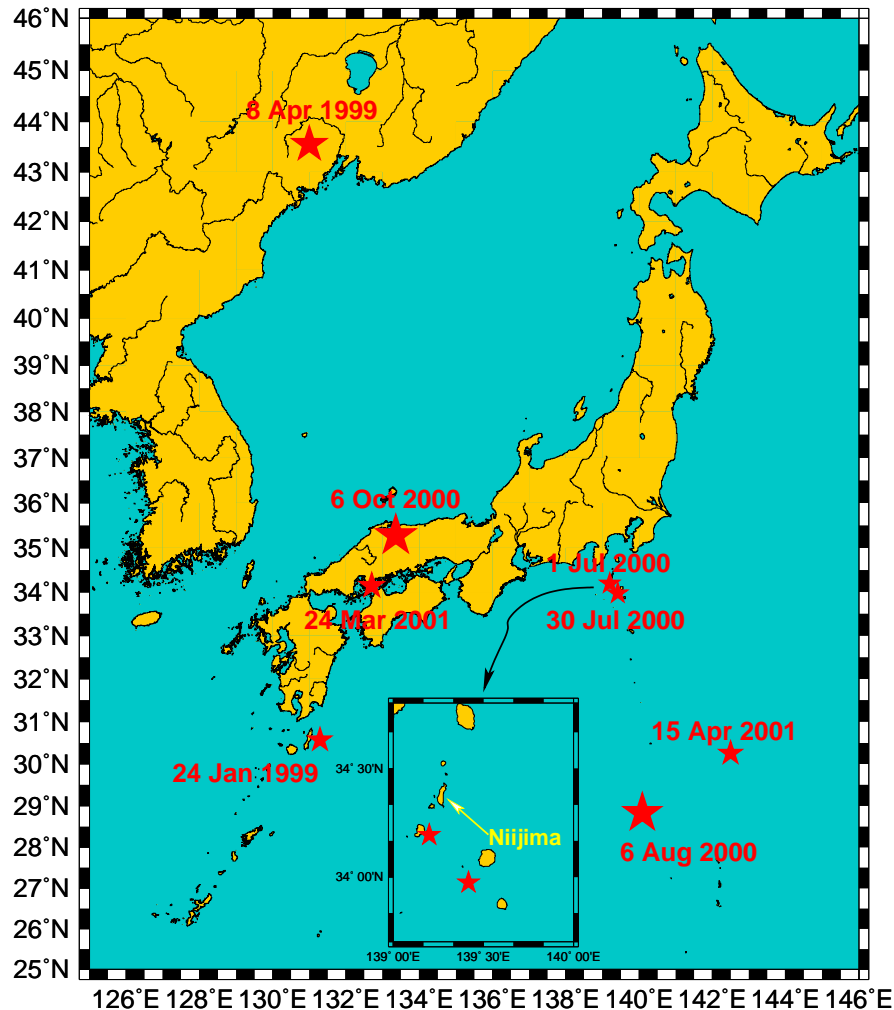}
\caption{(color online) The red stars show the epicenters of all
earthquakes reported by the JMA seismic catalog of magnitude
comparable to or larger than that of the earthquake on 1 July 2000
during the 3 year period from 1 January 1999 until 1 January 2002.
The inset depicts in an expanded scale the rectangular area
considered in the natural time analysis of seismicity by
\citet{UYE09}. Niijima Island at which their measuring station has
been installed is also shown.}\label{fig2}
\end{figure}

Figure \ref{fig2} shows the epicenters (red stars) within the area
25-46$^o$N 125-146$^o$E of all EQs of magnitude comparable to or
larger than that of the EQ on 1 July 2000 for a three year period
from 1 January 1999 until 1 January 2002 extending from $1
\frac{1}{2}$ year before until $1 \frac{1}{2}$ year after the case
discussed here. An inspection of this map reveals that eight EQs
occurred, six of which had epicenters several hundreds kilometers
away from the Niijima Island measuring station and two in its
vicinity. Only the latter two EQs on 1 July 2000 and 30 July 2000
could have been preceded by SES activities recorded at Niijima
station, in view of the up to date observations \citep{SPRINGER}
that magnitude 7.0 class EQs give detectable SES at epicentral
distances up to around 250 km or so. In other words, the lack of
geoelectrical data from stations that would have been installed in
the regions surrounding the six distant EQs from Niijima Island,
imposes the following constraint in order to achieve the goal of
the present study: Only the analysis during the period preceding
the occurrence of the two EQs in the neighborhood of Niijima
Island (i.e., before 1 July 2000) could serve for the purpose of
our study. Furthermore, and in order to avoid any influence on our
results from the aftershock activity of the previous EQ that
occurred on 8 April 1999 at 130.99$^o$N 43.55$^o$E of magnitude
7.1, we started our analysis some months later, i.e., on 1 October
1999 until just before the occurrence of the EQ on 1 July 2000.
Nevertheless, despite this limitation imposed primarily from the
{lack of} geoelectrical data {at a
multitude of measuring stations}, we note the following privilege:
The aforementioned SES activity at Niijima Island we considered
here, is well isolated in time and space \citep[since][noticed
that, beyond the SES activity they reported, which started almost
two months before the onset of the swarm activity, no other SES
activities have been recorded at Niijima station either well
before the onset or after the cessation of the swarm activity,
e.g., see the caption of Fig. 9 of \citealt{UYE09}]{UYE02,UYE09}.
This provides further convincing evidence in favor of our main
finding and excludes any possibility of attributing it to chance,
as already mentioned in the main text.

{The following two comments are now in order:}

\begin{figure}
\includegraphics[scale=1.0]{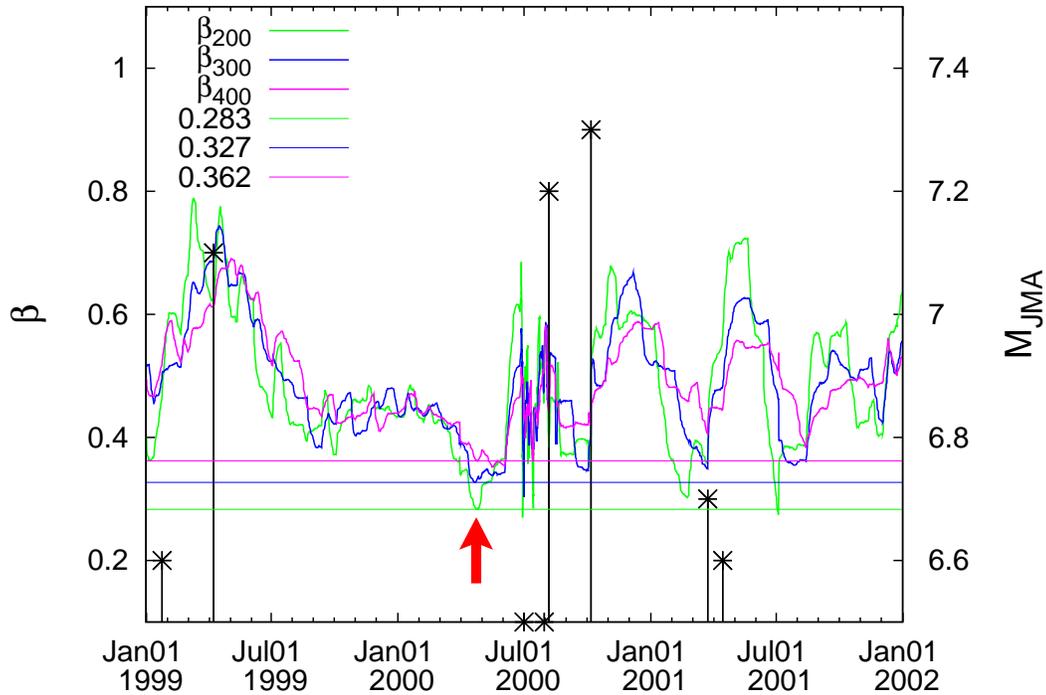}
\caption{{(color online) The same as Fig. \ref{fig1}, but extended
to a three year period from 1 January 2000 to 1 January 2002 for
$W$=200 (green), $W$=300 (blue) and $W$=400 (magenta). The thick
red arrow shows the minimum also marked in Fig. \ref{fig1}(b), the
values of which for $W$=200, 300 and 400 are designated by the
horizontal lines of the corresponding color. The EQs with
magnitudes $\rm {M_{JMA}}\geq6.5$ (right scale) are shown with
vertical bars ending at asterisks.}}\label{figA2}
\end{figure}

{First, the analysis of the variability $\beta$ of $\kappa_1$ of
seismicity versus the conventional time, the results of which were
presented as mentioned in Fig. \ref{fig1}(a) for the nine month
period 1 October 1999 to 1 July 2000, was extended further into
the past and the future. The relevant results during the three
year period from 1 January 1999 to 1 January 2002 are now depicted
in Fig. \ref{figA2} for the three longer natural time window
lengths $W$=200, 300 and 400 events (the fluctuations of which are
evidently appreciably smaller than those for $W$=100 -see Fig.
\ref{fig1}(a)- thus, if the latter were plotted in Fig.
\ref{figA2} it would overload this figure). For the sake of
reader's convenience, we also draw in Fig. \ref{figA2} three
horizontal lines that correspond to the minimum of the variability
$\beta$ of $\kappa_1$, around the date 26 April 2000 (shown here
by the thick red arrow), for $W$=200 (green) $W$=300 (blue) and
$W$=400 (magenta), respectively, already marked in Fig.
\ref{fig1}(b). We clarify that, as already mentioned, the
investigation of the interconnection of these minima with the
subsequent major EQs, including the criteria that distinguish
which of these minima are of truly precursory nature, was the
objective of a separate study, see \citet{PNAS13}, extended to an
almost 27 year period (1984-2011). The aim of the present paper
is{, as already mentioned,} essentially different:
By focusing solely on the periods during which a pronounced SES
activity has been recorded, as the one ($\sim$ 26 April 2000)
discussed here, to investigate whether there exists also an almost
simultaneous change in the variability $\beta$ of $\kappa_1$ of
seismicity that is statistically significant.}

\section*{{Acknowledgements}}
{We would like to express our sincere thanks to
Professor H. Eugene Stanley, Professor Shlomo Havlin and Dr. Joel
Tenenbaum for providing us the necessary data in order to insert  in Fig. 2
the nodes and the associated links of their network.}

\end{document}